\documentclass[preprint,12pt]{aastex}

\newcommand{\lya}{\mbox{Ly$\alpha$} }
\newcommand{\phialpha}{\mbox{$\varphi_{\alpha}\left(\lambda ,\Omega \right)$} }
\newcommand{\phialphar}{\mbox{$\varphi_{\alpha}^\mathrm{r}\left(\lambda
\right)$} } 
\newcommand{\phialphaz}{\mbox{$\varphi_{\alpha}^\mathrm{z}\left(\lambda
\right)$} } 
\newcommand{\phialphamean}{\mbox{$\varphi_{\alpha}^\mathrm{mean}\left(\lambda
\right)$} }
\newcommand{\hone}{\ion{H}{1} }
\newcommand{\htwo}{\ion{H}{2} }
\newcommand{\lumir}{\mbox{$\tilde{L}_{\alpha}^\mathrm{r}$} }
\newcommand{\lumiz}{\mbox{$\tilde{L}_{\alpha}^\mathrm{z}$} }
\newcommand{\lumimean}{\mbox{$\tilde{L}_{\alpha}^\mathrm{mean}$} }

\shortauthors{
M.A.R.Kobayashi,
Kamaya,
\& Yonehara
}

\shorttitle{
LYMAN ALPHA LINE SPECTRA OF THE FIRST GALAXIES
}

\begin{document}

\title{
LYMAN ALPHA LINE SPECTRA OF THE FIRST GALAXIES: DEPENDENCE ON OBSERVED
DIRECTION TO THE UNDERLYING CDM FILAMENT
}

\author{
Masakazu A.R. Kobayashi\altaffilmark{1},
Hideyuki Kamaya\altaffilmark{1},
and Atsunori Yonehara\altaffilmark{2,3}
}

\altaffiltext{1}{
Department of Astronomy, School of Science, Kyoto University,
Sakyo-ku, Kyoto 606-8502, JAPAN
}

\altaffiltext{2}{
Department of Physics, The University of Tokyo, Bunkyo-ku, 
Tokyo 113-0033, JAPAN
}

\altaffiltext{3}{
Inoue Fellow
}

\email{
kobayasi@kusastro.kyoto-u.ac.jp
}

%%%%%%%%%%%%%%%%%%%%%%%%%%%%%%%%%%%%%%%%%%%%%%%%%%%%%%%%%%%
% The abstract should be a single paragraph of            %
%  not more than 250 words, and should not contain        %
%  reference citations.                                   %
%%%%%%%%%%%%%%%%%%%%%%%%%%%%%%%%%%%%%%%%%%%%%%%%%%%%%%%%%%%
\begin{abstract}

 The first galaxies in the Universe are built up where cold dark matter
 (CDM) forms large scale filamentary structure. Although the galaxies
 are expected to emit numerous \lya photons, they are  surrounded by
 plentiful neutral hydrogen with a typical optical depth for \lya of
 $\sim 10^5$ (\hone halos) before the era of cosmological
 reionization. The \hone halo almost follows the cosmological Hubble
 expansion with some  anisotropic corrections around the galaxy because
 of the gravitational attraction by the underlying CDM filament. In this
 paper, we investigate the detectability of the \lya emissions from the
 first galaxies, examining their dependence on viewing angles. Solving
 the \lya line transfer problem in an anisotropically expanding \hone
 halo, we show that the escape probability from the \hone halo is the
 largest in direction along the filament axis. If the \lya source is
 observed with a narrow-band filter, the difference of apparent \lya
 line luminosities among viewing angles can be a factor of $\gtrsim 40$
 at an extreme case. Furthermore, we evaluate the predicted physical
 features of the \lya sources and flux magnification by gravitational
 lensing effect due to clusters of galaxies along the filament. We
 conclude that, by using the next generation space telescopes like the
 JWST, the \lya emissions from the first galaxies whose CDM filament
 axes almost face to us can be detected with the S/N of $\gtrsim 10$.

\end{abstract}

%%%%%%%%%%%%%%%%%%%%%%%%%%%%%%%%%%%%%%%%%%%%%%%%%%%%%%%%%%%%
% The subject headings (a maximum of six) should be listed %
%  after the abstract.                                     %
% The current list of subject headings is printed in       %
%  the Annual Index to the Journal and is available online %
%  at http://www.journals.uchicago.edu/ApJ/keywords.html.  %
%%%%%%%%%%%%%%%%%%%%%%%%%%%%%%%%%%%%%%%%%%%%%%%%%%%%%%%%%%%%

%%%%%%%%%%%%%%%%%%%%%%%%%%%%%%%%%%%%%%%%%%%%%%%%%%%%%%%%%%%%%%%%%%%%%
\section{INTRODUCTION}

One of the most important questions in cosmology is how galaxies are
formed and evolved in the context of the cold dark matter (CDM)
Universe. To answer the question, it is essential to search for young
galaxies at high-redshift systematically and to study their
observational properties in detail. In the hierarchical structure
formation scenario of the CDM Universe, the first bright objects and/or
the first galaxies are predicted to be formed in about some hundreds
Myrs after the Big Bang (i.e., redshift $\sim 10$) and to be embedded in
the deep gravitational potential wells of the CDM \citep[see
e.g.,][]{abel98, bert98, bcl99, yoshida03}. They are expected to be
luminous at the \lya line \citep{pp67} because the strong ionizing
radiation from young massive stars in the galaxies should lead to
prominent \lya emission through the recombination of hydrogen in their
interstellar medium.

In this decade, many galaxies which appear to be at early stages have
been observed \citep[e.g.,][]{rhoads00, steidel00, ajiki03, shapley03,
hayashino04, matsuda04, wang05}. Recently, the galaxies beyond redshift
6 have been frequently detected by their prominent \lya emission with
faint continuum (\lya emitters [LAEs]) by using ground-based large
telescopes like Subaru \citep{kodaira03, nagao04, taniguchi05,
shioya05}, Keck \citep{hu02, rhoads04} and VLT \citep{cuby03, kurk04,
tran04}. Moreover, surprisingly, the possible detection of a redshift
$\sim 10$ galaxy is reported by \citet{pello04}, although this is still
controversial \citep{bremer04, weatherley04}. In the following decade,
next generation space telescopes like the \textit{James Webb Space
Telescope} (JWST) are launched. The expected limiting $10\sigma$ flux of
JWST for a point source reaches to nJy-level ($\sim 10^{-31}\
\mathrm{erg\ s^{-1}\ cm^{-2}\ Hz^{-1}}$) in the wavelength range of
$\sim 1-5\ \micron$; this range corresponds to the redshifted \lya
wavelength at the source redshift of $7-40$. Thus, it seems to be in
not-so-distant future that the \lya emission from the first galaxies
and/or the LAEs beyond redshift of 10 are firmly detected.

However, before the era of cosmological reionization, these first
galaxies should be surrounded by plentiful neutral hydrogen in the
intergalactic medium (IGM); we call this ``\hone halo'' from now on. For
completely neutral Universe, typical scattering optical depth for the
\lya photons escaped from the galaxies at redshift about 10 is estimated
to be $\sim 10^5$ by using the standard cosmological parameters. One
might be discouraged by the extremely high optical depth for the \lya
photons, but it has been shown that most of them can escape from the
\hone halo around a \lya source at young Universe and travel freely
toward the observer \citep{lr99}. The reasons are as follows: (1) \lya
photons are not destroyed in the intergalactic space because
free-electrons and dust as sources of the two-photon decay and the
ultraviolet continuum absorption, respectively, rarely exist in the IGM
in such early Universe; (2) in the \hone halo, which is assumed to
expand following the pure cosmological Hubble flow, the \lya photons are
scattered many times and diffused redward in wavelength; at last,
cumulative scattering redshift grows large enough to escape from the
\hone halo.

\begin{figure}
% \figurenum{1}
% \epsscale{0.5}
 \plotone{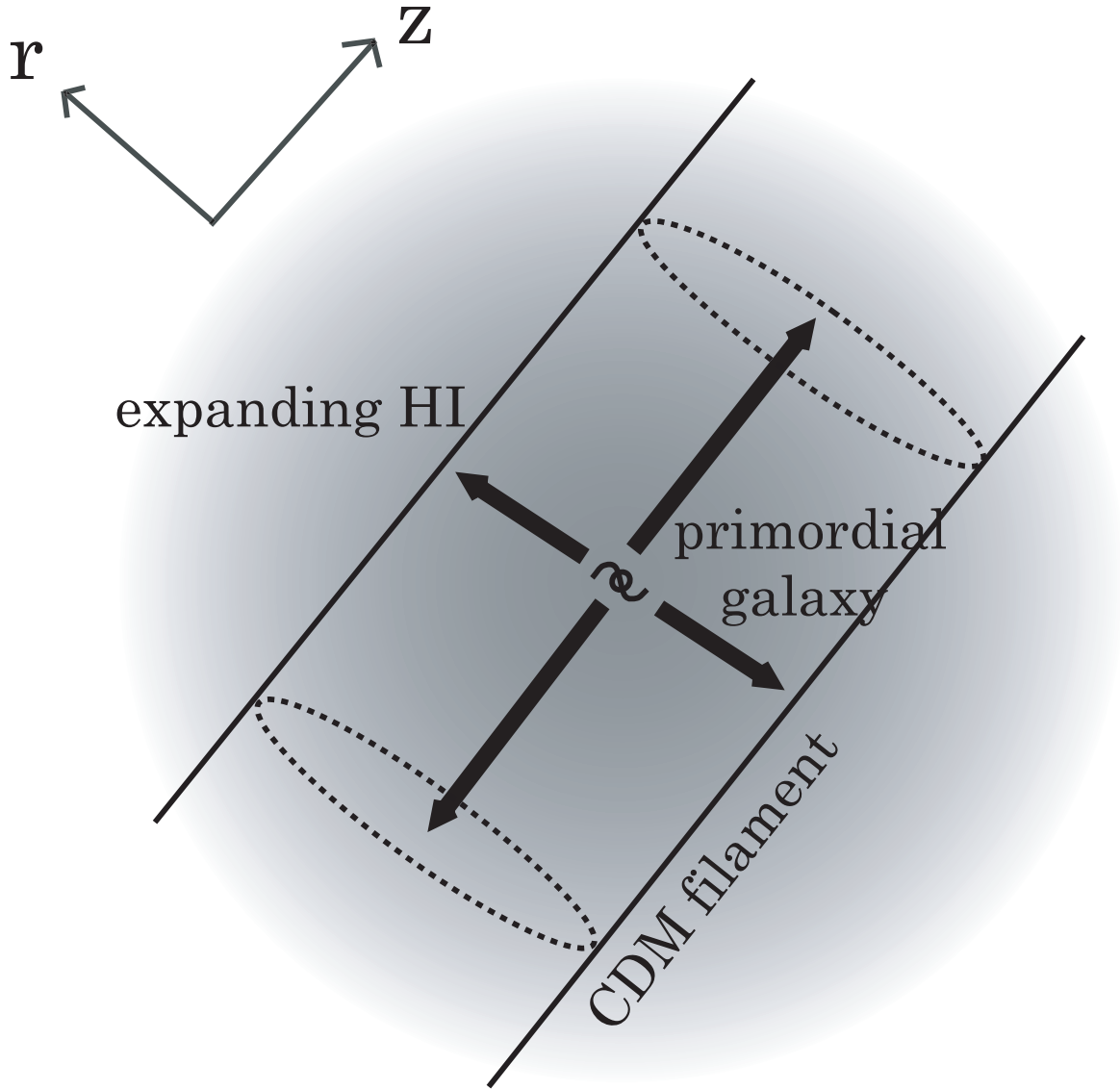}
 \caption{
 Schematic illustration of one of the first galaxies and the underlying
 CDM filament surrounded by an anisotropically expanding \hone halo with
 uniform density distribution. $z$- and $r$-axis (all directions
 perpendicular to $z$-axis) are also depicted. The thick arrows around
 the galaxy represent the expanding speed of \hone medium assumed in
 this paper; along $r$-axis, the \hone halo decelerated to some extent
 compared to that along $z$-axis (pure Hubble expansion) owing to the
 gravitational attraction force of the CDM filament.
 }
 \label{fig1}
\end{figure}

The hierarchical structure formation scenario of the CDM Universe
indicates that the first galaxies are formed at high-density peak
regions in the CDM distribution. In such regions, the CDM structures are
predicted to be filamentary by numerical N-body simulations
\citep[e.g.,][]{yoshida03}. The underlying CDM filament may affect the
\hone halos around them by gravitational attraction force. It is
naturally expected that the gravity works anisotropically; it is the
strongest in direction perpendicular to the axis of the CDM filament
($r$-direction), while it is the weakest along the axis direction
($z$-direction).
%-- Figure 1 should be placed here --%
%\placefigure{fig1}
%-- Figure 1 should be placed here --%
This situation around one of the first galaxies is illustrated
schematically in figure \ref{fig1}. Considering such situation, we
assume an improved expansion law to the \hone halo in our previous paper
\citep[KK04 in short]{kk04} (\textit{spheroidal expansion law}; see
equation (\ref{eq1}) below): the \hone halo expands following a pure
Hubble expansion in $z$-direction, while it is decelerated to some
extent from the pure Hubble flow in $r$-direction.

We have investigated the effects of the anisotropic expansion of the
\hone halo to the \lya line profiles and luminosities of the first
galaxies in KK04. We have found the following two effects: (1) the peak
of the \lya line is redshifted to longer wavelength; (2) the full width
at half maximum (FWHM) of the line is broadened to wider than those of
the isotropic expansion model to the \hone halo. Since the FWHM
corresponds to the Doppler width at the temperature of $\sim 10^8\
\mathrm{K}$, it has been predicted that no intrinsic \lya line features
of the first galaxies appear. As deceleration against the Hubble
expansion by the gravity of the filament becomes strong, these two
effects result in less luminosity of the diffuse \lya emission line.

Furthermore, as expected from the anisotropy in the expansion of the
\hone halo, the \lya line profile has dependence on the direction from
which the \lya source galaxy is observed. Clarifying such dependence is
essential to discuss the detectability of \lya emission lines from the
first galaxies. The dependence can be studied by dividing the solid
angle around the source into some regions and by investigating the \lya
line profile in each region. However, we have not been able to examine
the dependence in KK04 to keep the high statistical reliance of the
numerical solution in a Monte Carlo simulation. This is because the
following reason: in a Monte Carlo simulation, the convergence of the
solution needs to be checked by increasing the number of test photons,
but the total number of test photons has been restricted by the weak
computational power to $\lesssim 10^8$ in KK04. We have improved the
numerical code to solve the transfer problem with larger number of test
photons than $\sim 10^8$. Therefore, this is the time to revisit the
\lya line transfer problem in the anisotropically expanding \hone halo
with the improved numerical code.

In this paper, we reexamine the \lya line transfer problem with 10 times
larger number of test photons than that of KK04. In \S2, followed by
KK04, we summarize the formalism of the \lya line transfer problem in an
expanding \hone halo and discuss about the model of the \hone halo. Our
numerical results such as \lya line profiles, apparent \lya line
luminosities, and typical angular sizes of the first galaxies in \lya
emission observed from the $z$- and $r$-directions are shown in
\S3. Then, in \S4, we give the \lya flux of the source with intrinsic
\lya line luminosity of $10^{43}\ \mathrm{ergs\ s^{-1}}$, and discuss
the expected signal-to-noise ratio (S/N) for the source in the case of
the observation by the JWST. Finally, we conclude the paper in \S5. We
adopt the \textit{WMAP} cosmological parameters in the whole
calculations of this paper, that is, $\Omega_b=0.044,\ \Omega_M=0.27,\
\Omega_{\Lambda}=0.73$ and $h_0=0.71$ \citep{spergel03}.

%%%%%%%%%%%%%%%%%%%%%%%%%%%%%%%%%%%%%%%%%%%%%%%%%%%%%%%%%%%%%%%%%%%%%
\section{FORMULATION AND CALCULATION}

As briefly described in introduction, we have formulated a \lya line
transfer problem in KK04 according to \citet{lr99}, setting a simple
assumption to the expansion law of \hone halo, that is,
\textit{spheroidal expansion law}:
\begin{equation}
 v = H_s\left[\varepsilon^2 \left(x^2+y^2\right)+z^2\right]^{1/2},
 \label{eq1}
\end{equation}
where $v$ is recession velocity, $H_s$ is the Hubble expansion rate at
the \lya photon source redshift of $z_s$, and $\varepsilon$ is the
asphericity or the decreasing rate of the recession velocity in
$r$-direction ($x-y$ plane) to the pure Hubble expansion. In other
words, this parameter $\varepsilon$ represents the relative intensity of
the gravitational attraction force by the underlying CDM filament to the
cosmological Hubble expansion. We choose the parameter of $\varepsilon$
as a free parameter from 1.0 (corresponds to the pure Hubble expansion
in all directions) to 0.3 with a step of 0.1. As noted in KK04,
$\varepsilon$ is assumed to be a single value globally for simplicity,
although it should have different values depending on the distance from
the CDM filament and approach to unity as leaving the galaxy far
behind. This assumption can be hold because of the following reason. The
properties of escaping \lya photons are determined essentially at a
position where the optical depth for the \lya photons decreases to about
unity and where the physical distance from the source galaxy is $\sim
r_{\ast}$. The characteristic proper distance of $r_{\ast}$ is a
function of $\varepsilon$ and gradually increases from 1.0 to 2.9 Mpc as
$\varepsilon$ decreases from 1.0 to 0.3. The gravity of the underlying
CDM filament can affect this region even at the early Universe
\citep{bkp96, ckc04}.

Let $I = I(\nu$, \mbox{\boldmath $\Omega$}, \mbox{\boldmath $r$}) be
the specific intensity of a \lya line (in photons $\mathrm{cm^{-2}\
s^{-1}\ sr^{-1}\ Hz^{-1}}$), where $\nu$ is comoving frequency,
{\boldmath $r$} is radial vector from the source galaxy, and {\boldmath
$\Omega$} is direction cosine vector. The comoving transfer equation for
a \lya line with the assumption of isotropic coherent scattering is
given by
\begin{equation}
 {\mbox{\boldmath $\Omega$}} \cdot \nabla I 
  +\alpha \psi \left(\varepsilon, w\right)\frac{\partial I}{\partial \nu}
  =\chi_{\nu}\left(J- I\right) + S.
  \label{eq-lya_trans}
\end{equation}
Here, $\nu$ is the redshift of the frequency $\nu_{\alpha}-
\nu_\mathrm{photon}$ where $\nu_{\alpha}$ is the resonant frequency of
\lya and $\nu_\mathrm{photon}$ is the photon frequency; $\chi_{\nu}$ is
the \lya scattering opacity at the frequency redshift $\nu$ (the
analytical form is represented in KK04); $J=(1/2)\int_{-1}^1Id\mu$ is
the mean intensity; $S$ is the \lya emission function; $\alpha \equiv
H_s\nu_{\alpha}/c$ is the increasing rate of the frequency redshift per
unit distance in the $z$-direction (i.e. the direction in which the
\hone medium in a \hone halo follows the pure Hubble expansion); and
\begin{equation}
 \psi \left(\varepsilon, w\right)=
  \left[\varepsilon^2+\left(1-\varepsilon^2\right)w^2\right]^{1/2},
  \label{eq-def_psi}
\end{equation}
represents the effect of the anisotropic expansion of a \hone halo where
$w$ is the $z$-axis component of direction cosine. The source function
on the right-hand side of equation (\ref{eq-lya_trans}) can be written
as
\begin{equation}
 S=\frac{\dot{N}_{\alpha}}{\left(4\pi \right)^2r^2}
  \delta{(\nu)}\delta{(r)},
\end{equation}
where $\dot{N}_{\alpha}$ is the steady emission rate of \lya photons by
the source (in photons $\mathrm{s}^{-1}$) and assumed to be constant.

In the whole calculation in this paper, we assume that a steady point
source galaxy at redshift 10 which lit up before the cosmological
reionization is surrounded by a uniform, completely neutral IGM. In
other words, we assume that the reionization epoch is redshift $\sim
10$, which is intermediate between the late epoch of $\sim 6$
\citep{becker01} and the early epoch of $\sim 20$
\citep{kogut03}. Apparently, the assumption of completely neutral IGM at
redshift $\sim 10$ is oversimplified since a local cosmological \htwo
region can spread over the source of \lya photons
\citep[e.g.,][]{haiman02, santos04a}. Although the so-called ``proximity
effect'' is important for quasars but is less so for galaxies, the
typical sizes of the \htwo regions around \lya source galaxies are
expected to be not much less than $\sim 1$ proper Mpc, which is
comparable to the size of the scattering region discussing in this paper
\citep{bl01}. So there simply must be a phase when \htwo regions around
early galaxies are becoming larger than the surrounding filaments; our
results can hold prior to the phase. Fortunately, as shown in KK04, the
\lya line profiles of the first galaxies at redshift $> 10$, which are
considered to be in such phase, are analogous with those of sources at
redshift 10 shown in this paper. However, some corrections for the model
assumed in this paper due to the \htwo regions around \lya sources and
the density fluctuations in \hone halo may be necessary.

Our calculation algorithm is identical with that of KK04. Avoiding
redundancy, we only summarize the Monte-Carlo technique used in the
calculations here (see \S 2 and \S 3 of KK04 for the details). At first,
according to an initial frequency redshift of test photons from the \lya
line center, the initial distances of them from the source are
determined stochastically by means of the diffusion solution of
\citet{lr99}. The tentative optical depth $\tau$ for a test photon is
determined as $-\ln{R_\mathrm{sca}}$, where $R_\mathrm{sca}$ is a random
number between 0 and 1, and the photon proceeds the distance
proportional to $\tau$. Once $\tau >\tau_\mathrm{max}\equiv
\int^{\infty}_{0} \chi_{\nu+\alpha \psi l}dl$, we say that the photon
escapes from the \hone halo. During scattering events, the frequency
redshift of a photon is increased according to the value of $\psi
\left(\varepsilon, w\right)$. These procedures are repeated until all
test photons escape from the \hone halo.

The total number of the test photons $N_\mathrm{tot}$ is $10^9$, which
is ten times larger than that of KK04. Thanks to the increase of
calculating photon number and the improvement of our numerical code, we
can divide the whole solid angle into $n$ regions with almost the same
statistical precision as that of KK04. These regions are selected to
have the same solid angle of $4\pi /n$ and to be axi-symmetric around
the filament axis. Thus, it can be discussed meaningfully how much the
\lya line profiles and the apparent \lya line luminosities of the first
galaxies depend on the direction from which they are observed. We adopt
$z_s=10$ and $n=10$ for all calculations shown in this paper.

%%%%%%%%%%%%%%%%%%%%%%%%%%%%%%%%%%%%%%%%%%%%%%%%%%%%%%%%%%%%%%%%%%%%%
 \section{RESULTS} 

First, we define the \lya line profile per unit wavelength per unit
solid angle (in $\mathrm{sr^{-1}\ \micron^{-1}}$),
$\varphi_{\alpha}\left(\lambda ,\Omega \right)$, as the following
equation:
\begin{equation}
 \varphi_{\alpha}\left(\lambda ,\Omega \right)d\lambda d\Omega 
  \equiv 
  \frac{N\left(\lambda ,\Omega \right)d\lambda d\Omega 
  \cdot \left[hc/\lambda \right]}
  {N_\mathrm{tot}\cdot \left[hc/\left(1+z_s\right)
			\lambda_{\alpha}\right]},
  \label{eq-def_tilL}
\end{equation}
where $N\left(\lambda ,\Omega \right)d\lambda d\Omega$ is the number of
escaping photons with the observed wavelength and solid angle
in ranges of $\left(\lambda ,\lambda +d\lambda \right)$ and
$\left(\Omega,\Omega+d\Omega\right)$, respectively;
$\lambda_{\alpha}=0.1216\ \micron$ is the resonant wavelength of \lya in
the rest-frame. The characteristic quantity on the denominator of the
right-hand side of equation (\ref{eq-def_tilL}) represents the initial
total energy of the test photons in the observer frame. Therefore, the
integrated value of \phialpha over the whole wavelength and solid angle
becomes unity only if the wavelength redshifts caused by scattering in
expanding \hone halos are negligible.

  \subsection{Mean line profiles and scattered line luminosities}

Before we investigate the dependence of the observed \lya line profile
and apparent \lya line luminosity of the first galaxies on viewing
angles to them, we show the characteristic effects of the anisotropic
expansion of a \hone halo to the \lya emission line according to
KK04. For this purpose, here we define two tentative quantities:
\textit{mean \lya line profile} and \textit{scattered \lya line
luminosity}, denoted by \phialphamean and
$\tilde{L}_{\alpha}^\mathrm{mean}$, respectively. We emphasize here that
they are not observational quantities except for those with $\varepsilon
=1$ but make the effects of the anisotropic expansion of a \hone halo
clear.

\begin{figure}
% \figurenum{2}
% \epsscale{0.5}
 \plotone{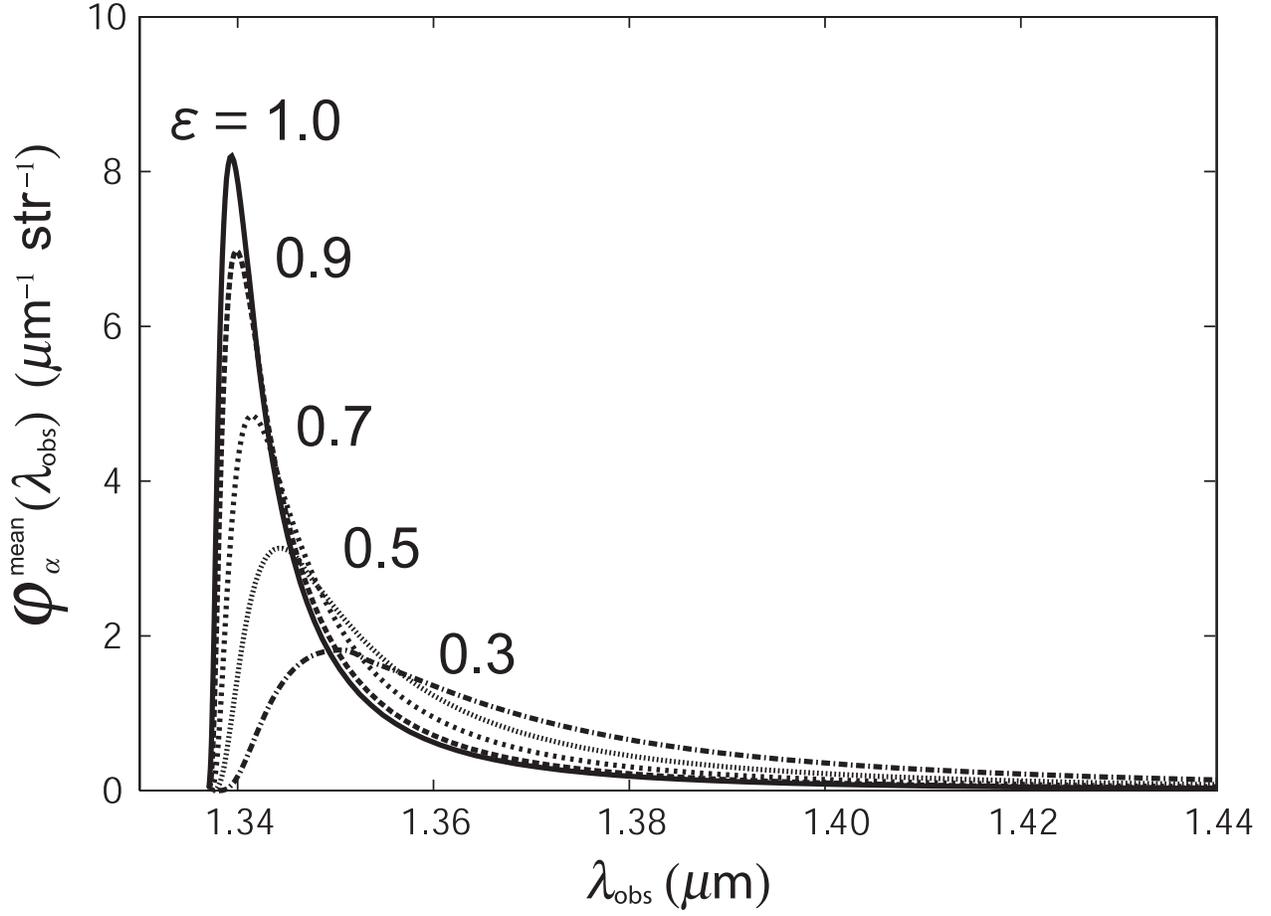}
 \caption{
 Mean line profiles of the scattered \lya line of one of the first
 galaxies at $z_s=10$, $\varphi_{\alpha}^\mathrm{mean}\left(\lambda
 \right)$, as a function of the observed wavelength
 $\lambda_\mathrm{obs}$ (in \micron) for various parameters of
 $\varepsilon$. They show the line profiles of $\varepsilon = 1.0$
 (\textit{solid line}; corresponds to the pure Hubble expansion), $0.9$
 (\textit{long dashed line}), $0.7$ (\textit{short dashed line}), $0.5$
 (\textit{dotted line}), $0.3$ (\textit{long dash-dotted line}) as
 labeled.
 }
 \label{fig2}
\end{figure}

For a certain $\varepsilon$, \phialphamean is obtained by counting the
number of test photons with the observed wavelength in range of
$\left(\lambda, \lambda+d\lambda \right)$ without distinction among
escape directions, $N(\lambda)d\lambda$:
\begin{equation} 
\varphi_{\alpha}^\mathrm{mean}\left(\lambda \right)
 =\frac{\left[N(\lambda)/4\pi \right]\cdot \left[hc/\lambda \right]}
 {N_\mathrm{tot}\cdot \left[hc/\left(1+z_s\right)\lambda \right]}.
\end{equation}
Note that \phialphamean is obtainable without knowing each line profile
which is observed from a viewing angle.
%-- Figure 2 should be placed here --%
%\placefigure{fig2}
%-- Figure 2 should be placed here --%
Figure \ref{fig2} presents the mean \lya line profiles for the various
values of the parameter $\varepsilon$. The solid, long-dashed,
short-dashed, dotted, long dash-dotted lines are $\varepsilon =1.0$ (the
Hubble expansion), 0.9, 0.7, 0.5, and 0.3, respectively. As
$\varepsilon$ decreases, the wavelength at which \phialphamean has its
peak value, $\lambda_\mathrm{peak}$, becomes longer as well as the peak
value itself becomes lower; as a result, \phialphamean becomes more
diffuse in wavelength. These lines are well in agreement with the
results of KK04.

The scattered \lya line luminosity normalized by the total energy of the
test photons in the observer frame, $\tilde{L}_{\alpha}^\mathrm{mean}$,
is obtained by the product of the integral of \phialphamean over the
whole wavelength and the whole solid angle $4\pi$:
\begin{equation}
 \tilde{L}_{\alpha}^\mathrm{mean}\equiv 
  4\pi \int^{\lambda_u}_{\lambda_l} \varphi_{\alpha}^\mathrm{mean}
  (\lambda)d\lambda .
  \label{eq-def_lum}
\end{equation}
This represents the total energy fraction of the scattered \lya line to
the intrinsic one and therefore the value of \lumimean is always lower
than (or perhaps equal to) unity. Recall that this is not an
observational quantity for that with $\varepsilon \neq 1$. $\lambda_l$
and $\lambda_u$ are selected to be $1.337\ \micron$
($=(1+z_s)\lambda_{\alpha}$ where $z_s=10$) and $2.00\ \micron$,
respectively; this wavelength range covers almost all photons escaped
from the \hone halo.
%-- Table 1 should be placed here --%
%\placetable{tab1}
%-- Table 1 should be placed here --%
The results of \lumimean are listed at the 7th column in table
\ref{tab1} for some $\varepsilon$. One can see that \lumimean is always
lower than unity and slightly decreases as $\varepsilon$ decreases.

  \subsection{Characteristic line profiles and apparent line
  luminosities}

\begin{figure}
% \figurenum{3}
% \epsscale{0.5}
 \plotone{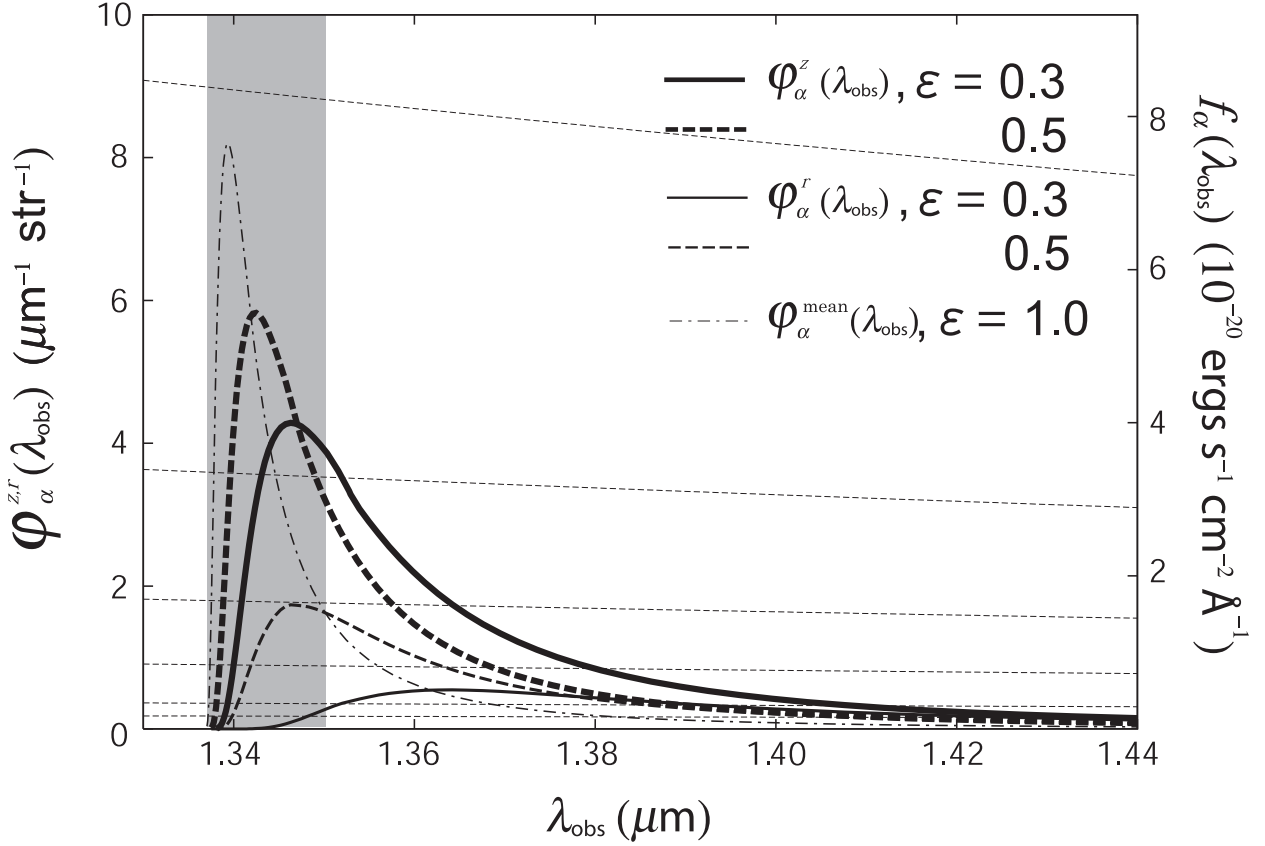}
 \caption{
 Some of the predicted \lya line profiles of the first galaxies at
 $z_s=10$. Two characteristic cases are depicted; \phialphaz and
 $\varphi_{\alpha}^\mathrm{r}\left(\lambda \right)$,  which are the
 profiles of the \lya source observed from $z$- and $r$-directions,
 respectively. The two solid lines and two dashed lines are $\varepsilon
 =0.3$ and $0.5$, respectively. The two thick lines with higher peak
 values in each kind of lines are
 $\varphi_{\alpha}^\mathrm{z}\left(\lambda \right)$, and the other two
 thin lines are $\varphi_{\alpha}^\mathrm{r}\left(\lambda
 \right)$. \phialphamean for $\varepsilon =1.0$ (\textit{dash-dotted
 line}) is also presented for reference.  The measure of observed flux
 densities of the \lya line, $f_{\alpha}(\lambda_\mathrm{obs})$, are
 also shown on the right-hand side of the vertical axis (in $10^{-20}\
 \mathrm{ergs\ s^{-1}\ cm^{-2}\ \AA^{-1}}$) with an intrinsic \lya line
 luminosity of $L_{\alpha}=10^{43}\ \mathrm{ergs\ s^{-1}}$. Contours
 show the expected JWST limiting $10\sigma$ flux for a point source with
 intrinsic \lya line luminosity of $L_{\alpha}=2.0\times 10^{40},
 5.0\times 10^{40}, 10^{41}, 2.0\times 10^{41}, 5.0\times 10^{41},
 10^{42}\ \mathrm{ergs\ s^{-1}}$ (from top to bottom). The shadowed
 region represents the bandpass of a tentative narrow-band filter
 adopted in this paper ($130\ \mathrm{\AA}$ centered at 1.344 \micron).
 }
 \label{fig3}
\end{figure}

Here we present the numerical results of \textit{observed \lya line
profiles} and \textit{apparent \lya line luminosities} of the first
galaxies in order to investigate the dependence of them on viewing
angles. First, we consider the observed line profiles of two
characteristic cases; one can be seen from $z$-direction and the other
from $r$-direction denoted by \phialphaz and
$\varphi_{\alpha}^\mathrm{r}\left(\lambda \right)$, respectively. These
are obtained by integrating \phialpha over the representative solid
angle of $4\pi/n$, and then, by multiplying the integrated value by a
factor of $n/4\pi$. We note here that, for $\varepsilon =1.0$,
\phialphaz and \phialphar completely correspond with each other and with
$\varphi_{\alpha}^\mathrm{mean}\left(\lambda \right)$.
%-- Figure 3 should be placed here --%
%\placefigure{fig3}
%-- Figure 3 should be placed here --%
The results are depicted in figure \ref{fig3} for $\varepsilon=0.3$
(\textit{thick} and \textit{thin solid lines}) and $0.5$ (\textit{thick}
and \textit{thin dashed lines}); the thick and thin lines are \phialphaz
and $\varphi_{\alpha}^\mathrm{r} \left(\lambda \right)$,
respectively. \phialphamean for $\varepsilon =1.0$ is also shown
(\textit{dash-dotted line}) for reference. One can see that, at the
whole range of the observed wavelength, \phialphaz with a certain
$\varepsilon$ has always larger value than that of \phialphar with the
same $\varepsilon$. In addition, \phialphar rapidly fades down as
$\varepsilon$ decreases, while \phialphaz doesn't.

The characteristic quantities of the observed \lya line profiles (i.e.,
$\lambda_\mathrm{peak}$ and the FWHM) are summarized in table
\ref{tab1}. At the 2nd and 3rd columns, $\lambda_\mathrm{peak}$ of
\phialphaz and \phialphar with various parameters of $\varepsilon$ are
compiled, respectively. It is shown that $\lambda_\mathrm{peak}$ of
\phialphar is redshifted to longer wavelength than that of \phialphaz
with the same $\varepsilon$. The 4th and 5th columns present the FWHMs
of \phialphaz and $\varphi_{\alpha}^\mathrm{r} \left(\lambda \right)$,
respectively, and the 6th column shows the ratio of the FWHMs of
\phialphaz to $\varphi_{\alpha}^\mathrm{r}\left(\lambda \right)$. These
indicate that \phialphar is always wider than \phialphaz with the same
$\varepsilon$ and that the difference among them grows further as
$\varepsilon$ decreases. For convenience, we also give the measure of
the observed flux per unit wavelength, $f_{\alpha}(\lambda)$, on the
right-hand side of the vertical axis in figure \ref{fig3} (in $10^{-20}\
\mathrm{ergs\ s^{-1}\ cm^{-2}\ \AA^{-1}}$), which is defined by
\begin{equation}
 f_{\alpha}(\lambda)\equiv 4\pi \varphi_{\alpha}(\lambda)\cdot
  \frac{L_{\alpha}}{4\pi d_L^2(z_s)}.
  \label{eq-flux-density}
\end{equation}
Here, $d_L(z_s)$ is luminosity distance from the observer to the source
redshift $z_s$. For $z_s=10$ which we adopt in this paper as the
redshift of the \lya emission source, $d_L$ is $3.27\times 10^{29}$ cm
($\sim 1.06\times 10^2$ Gpc). We apply a typical \lya line luminosity of
high-redshift LAEs of $L_{\alpha}=10^{43}\ \mathrm{ergs\ s^{-1}}$
\citep[e.g.,][]{taniguchi03,hu04, mr04, santos04b} as a reference
luminosity of the \lya source at $z_s=10$; this is based on the fact
that the luminosity function of LAEs in the redshift from 3 to 6 does
not significantly change \citep{mr04,van05}. The expected JWST limiting
$10\sigma$ flux for a point source of 1 nJy to the source at redshift 10
with intrinsic \lya line luminosities of $L_{\alpha}=2.0\times 10^{40},
5.0\times 10^{40}, 10^{41}, 2.0\times 10^{41}, 5.0\times 10^{41},
10^{42}\ \mathrm{ergs\ s^{-1}}$ is also presented by contours in figure
\ref{fig3} (from top to bottom).

The apparent \lya line luminosities can be obtained by integrating the
observed line profile in a certain range of wavelength and by
multiplying the integrated value by a factor of $4\pi$ like \lumimean in
equation (\ref{eq-def_lum}). This multiplication of the whole solid
angle reflects that, when a \lya source at $z_s=10$ is observed, the
\lya line luminosity is estimated under the assumption that the \lya
emission from the source is isotropic with the observed flux
density. Thus, this apparent \lya line luminosity coincides with the
scattered \lya line luminosity \lumimean only in the case that the \lya
photons isotropically escape from the \hone halo. In this paper, we
adopt the integral interval from $\lambda_l=1.337\ \micron$ to
$\lambda_u = 1.350\ \micron$, assuming the source is observed with a
tentative narrow-band filter centered at $1.344\ \micron$ with a
bandwidth of $130\ \mathrm{\AA}$. Then, we can obtain the apparent \lya
line luminosities of \lumiz and \lumir for \phialphaz and
$\varphi_{\alpha}^\mathrm{r}(\lambda)$, respectively. \lumiz and \lumir
are listed at the 8th and 9th column in table \ref{tab1},  respectively,
and their ratio is given at the last column. Compared with \lumir at the
same $\varepsilon$, \lumiz is always brighter. As $\varepsilon$
decreases, \lumir rapidly decreases while \lumiz remains high value;
therefore, the ratio of \lumiz to \lumir increases and results in the
factor of $\ga 43$ for $\varepsilon \la 0.3$.

  \subsection{Surface brightness distributions}

\begin{figure}
% \figurenum{3}
% \epsscale{0.5}
 \plotone{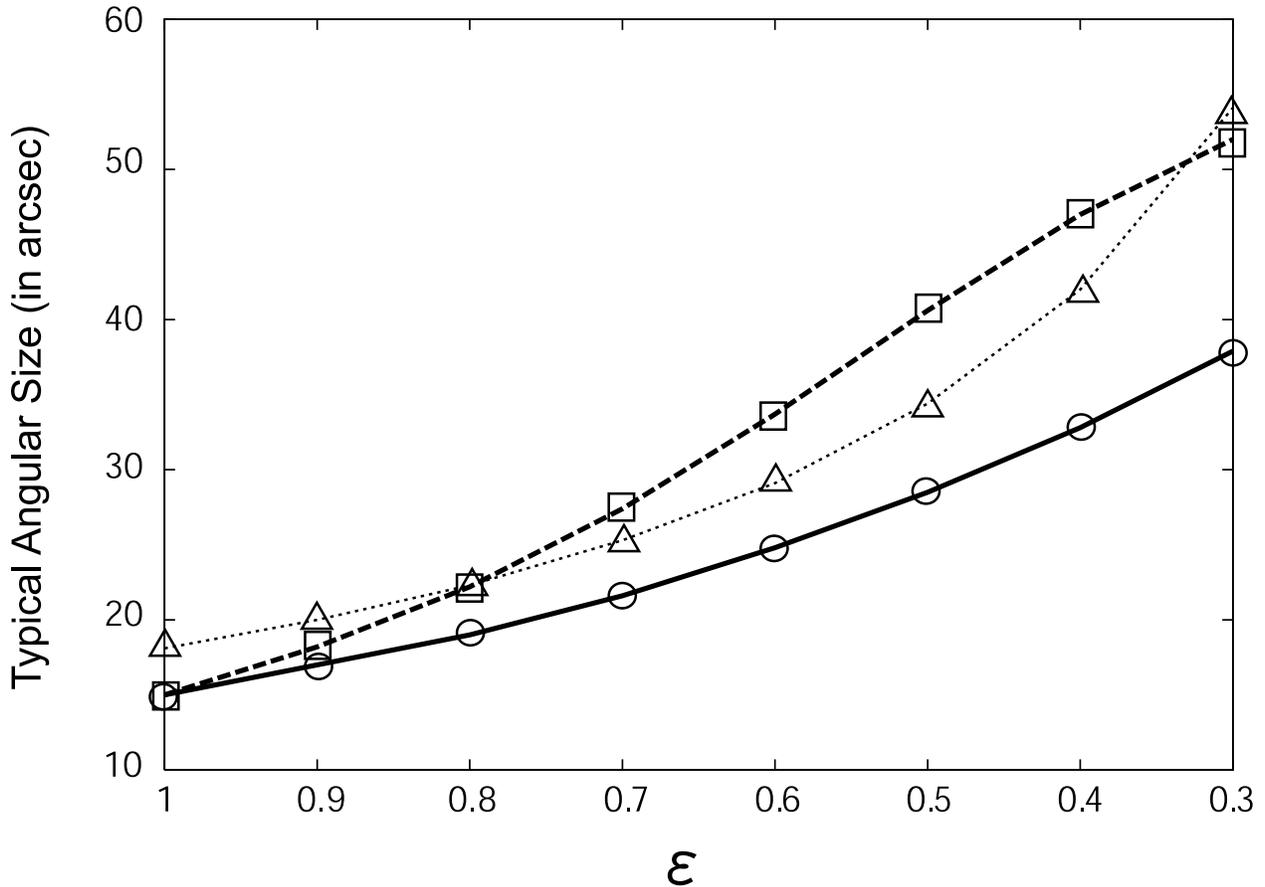}
 \caption{
 Typical angular radii of the first galaxies at \lya line on the sky (in
 arcsec) as functions of the parameter $\varepsilon$. The open circles
 with thick solid line and the open squares with thick dashed line are
 the typical angular sizes of the sources observed from the $z$- and
 $r$-directions, respectively. The open triangles with thin dotted line
 are the typical angular sizes of a characteristic distance from the
 source of $0.1r_{\ast}$ for a reference. Note that the open circles,
 open squares, and open triangles are the numerical results, while the
 various lines are the linear interpolators of them.
 }
 \label{fig4}
\end{figure}

Our results also indicate that \lya emissions from the first galaxies
are no longer point sources but diffuse on the sky; this agrees with the
result of the previous work by \citet{lr99}. This means that the surface
brightnesses of the first galaxies at the \lya line become low. As well
known, it is more difficult to detect more diffuse sources on the sky
with same total luminosity \citep{yoshii93}.
%-- Figure 4 should be placed here --%
%\placefigure{fig4}
%-- Figure 4 should be placed here --%
In order to examine whether the predicted \lya source can be detected by
the JWST, here we present the typical angular radius of the source at
\lya line, in which the number fraction of incoming \lya photons becomes
0.90, in figure \ref{fig4} for various $\varepsilon$. The open circles
with the thick solid line and open squares with the thick dashed line
represent the typical angular radii of the source observed from the $z$-
and $r$-directions, respectively. We also present the angular radius
corresponding to the physical radius of $0.1r_{\ast}$ as the open
triangles with the thin dotted line for reference. Numerical results are
the open circles, open squares, and open triangles, while the various
lines are the linear interpolators of them. Regardless of the observed
direction, the typical angular size grows monotonically as $\varepsilon$
decreases; it ranges from $15\arcsec$ to $\gtrsim 50\arcsec$. The
typical angular size of the source increases most rapidly in the case
that the source is observed from $r$-direction.

%%%%%%%%%%%%%%%%%%%%%%%%%%%%%%%%%%%%%%%%%%%%%%%%%%%%%%%%%%%%%%%%%%%%
\section{DISCUSSION}

According to the results described in the previous section, we discuss
the effects of the anisotropic expansion of a \hone halo to the observed
\lya line profile and the apparent \lya line luminosity of a redshift 10
galaxy, and then, examine the detectability of the \lya emission by the
JWST.

  \subsection{\boldmath Detectability of \lya emission: general
  comments}

As shown in figure \ref{fig2} and summarized at the 7th column in table
\ref{tab1}, the mean observed \lya line profile becomes more diffuse in
wavelength and the scattered \lya line luminosity becomes dimmer as
$\varepsilon$ decreases.
This result implies that, on average, the detectability of the first
galaxies by their \lya emissions gets worse as the expansion law of the
\hone halos becomes more anisotropic. Moreover, as $\varepsilon$
decreases, the discrepancy between the peak wavelength of the scattered
\lya line and the resonant wavelength of \lya in the observer frame
becomes more significant. That is, $\lambda_\mathrm{peak}$ of the \lya
line of the first galaxies does not correspond to
$\left(1+z_s\right)\lambda_{\alpha}$. This discrepancy causes a blunder
of the \lya source redshift. In order to estimate the source redshift
correctly, other lines of the same source such as H$\alpha$ need to be
detected.

However, once the object with a very broad emission line which is very
diffuse on the sky is detected, the effect of scattering in a \hone halo
can be easily recognized. This is because the FWHM of the \lya line is
anomalously broad and corresponds to a Doppler width at the temperature
of $\sim 10^8\ \mathrm{K}$; this is not easily attainable by any other
physical processes. Although the features of the \lya emissions from the
first galaxies have not been fully understood yet, our results will
present useful information because any intrinsic profiles can be
concealed by this modified line profile.

  \subsection{\boldmath Dependence of detectability of \lya emission on
  observed direction}

Turning now to the dependence of detectability of \lya emissions from
the first galaxies at redshift 10 on viewing angles, we find the
following anisotropic effects of the \hone halos. First, similar to the
mean \lya line, the observed \lya line profile in each viewing angle
becomes more diffuse in wavelength as $\varepsilon$ decreases; this
makes the apparent \lya line luminosity less luminous as $\varepsilon$
decreases. Second, however, the dimming and diffusing rates of the \lya
line depend on observed direction; the apparent \lya line luminosity
falls most sharply in $r$-direction, while it hardly decreases in the
$z$-direction as $\varepsilon$ decreases.

The differences of the \lya line profile and of the apparent \lya line
luminosity among viewing angles are understood by the dependence of the
optical depth for \lya photon on propagative direction due to the
anisotropic expansion of the \hone halo. That is, because of the
anisotropy in the expansion law of the \hone halo, the optical depth is
the largest for photons which propagate into the $r$-direction, while it
is the smallest for those which travel to the $z$-direction. This means
that escaping into $r$-direction is the most difficult; it needs to
experience more scattering events and to get larger cumulative
wavelength redshift by scattering compared to that into
$z$-direction. Therefore, we conclude that the direction along the axis
of the underlying CDM filament is the most profitable to detect the \lya
emission lines of the first galaxies.

In any case, the observed \lya line profiles and the apparent \lya line
luminosities of the first galaxies can be quite different from their
intrinsic profiles and luminosities. This modified \lya line profile
also needs to be fully understood in order to derive some physical
quantities of the first galaxies, e.g., star formation rates (SFRs)
and/or the escape fractions of the ionizing photons from their observed
\lya line profiles \citep[e.g.,][]{escude98,ch00,mr00,mhc04}.

The flux density represented by equation (\ref{eq-flux-density}) gives a
characteristic value as following:
\begin{equation}
 f_{\alpha}(\lambda)
  = 9.3\times 10^{-21}\left(\frac{L_{\alpha}}{10^{43}\ 
		       \mathrm{ergs\ s^{-1}}}\right)
  \left(\frac{\varphi_{\alpha}(\lambda)}{1.0\ \mathrm{\micron^{-1}\ 
   sr^{-1}}}\right)\ \mathrm{ergs\ s^{-1}\ cm^{-1}\ \AA^{-1}}.
\end{equation}
This typical value of $9.3\times 10^{-21}\ \mathrm{ergs\ s^{-1}\
cm^{-1}\ \AA^{-1}}$, which corresponds to $56\ \mathrm{nJy}$ and the AB
magnitude of 27.0 mag at $1.34\ \mathrm{\micron}$, is about one order of
magnitude brighter than the limiting $10\sigma$ flux of JWST for a point
source at the North Ecliptic Pole\footnote{For more details, see
http://www.stsci.edu/jwst/}. However, as shown in figure \ref{fig4}, our
results indicate that \lya emissions from the first galaxies are diffuse
on the sky; their typical angular sizes are a function of $\varepsilon$
and viewing angle, and range from $15\arcsec$ to $\sim 50\arcsec$.

Evaluating S/N for the predicted \lya emission with an intrinsic \lya
line luminosity of the source of $L_{\alpha}=10^{43}\ \mathrm{ergs\
s^{-1}}$ by utilizing the JWST Mission Simulator (JMS)\footnote{JMS can
be used at http://www.stsci.edu/jwst/science/jms/index.html}, we find
that the S/N for the source with $\varepsilon \leq 1.0$ is always
smaller than 10 over a $\gtrsim 100,000$ second exposure. Thus, it might
be an observational challenge to detect the \lya emission against the
very bright zodiacal light, even though it is at the North Ecliptic Pole
and observed from the direction along the filament axis at the
source. Furthermore, in the direction along the filament axis, there are
expected to be clusters of galaxies. The presence of them at the
foreground of the first galaxies might make the detection of the \lya
emissions more difficult because it is very difficult to discriminate an
object with low surface brightness against the one with higher surface
brightness at the foreground of it.

However, these clusters of galaxies can boost the detectability of the
background \lya emission by gravitational lensing. According to the
recent successful observations of the high-redshift LAEs by
gravitational lensing \citep{kneib96, ellis01, kneib04, pello04,
egami05}, high magnifications ($\sim 10-100$ times) can be
occurred. This high magnification is occurred only if the background
source is in the so-called ``critical regions''. Location of the region
is precisely known for well-understood clusters and depends on a
redshift range of the source \citep{kneib96, ellis01}. By using the
singular isothermal sphere (SIS) lens model \citep[e.g.,][]{tog84} with
the Einstein ring radius of $2\arcmin$ and Gaussian profiles as the
surface brightness of the source, it is shown that the predicted \lya
emission source can be magnified by a factor of $\gtrsim 10$. In this
case, the resultant S/N reaches $\ga 10$ over a $\ga 100,000$ second
exposure. Thus, we conclude that detection of the predicted \lya source
at the redshift $z_s\sim 10$ is feasible over this decade with the
JWST.

  \subsection{Implications of detecting \lya emission from the first
  galaxies}

We insist that the \lya line profiles of the sources at redshift $> 10$
are analogous with those of a source at redshift 10 presented in this
paper. They almost coincide with each other in dimensionless frequency
redshift space as shown in figure 6 of KK04. Therefore, in order to
evaluate the \lya line luminosities of the source at redshift $> 10$ and
the typical angular radius of it at the \lya line, it is sufficient to
know the characteristic frequency redshift and the characteristic proper
distance at each redshift given in KK04.

Furthermore, our results indicate that there is a correlation between
the apparent \lya line luminosity of the first galaxies and the
underlying CDM structures at the initial contraction stage. It is also
shown that this correlation is further enhanced if the \lya source is
observed by using a narrow-band filter targeting to a specific redshift
as the recent observations do. If one of the first galaxies at redshift
$\sim 10$ is detected, this correlation can give us useful information
that the underlying CDM filament at the extremely early Universe
(cosmological age is younger than $0.5$ Gyr) almost faces us, although
degeneracy with the intrinsic \lya line luminosity of the galaxy
remains.

Possibly, however, \lya sources at $z_s\ga 10$ are too faint to be
detected as single \lya emission sources, but rather the assembly of
them can be detected as a diffuse background source at near-infrared
wavelength. This may have already been detected as the cosmic
near-infrared background (CNIB), reported by independent groups
\citep{wr00, crbj01, matsumoto04} and cannot be explained by normal
galaxy populations \citep{totani01}. Thus, considered with our results,
the CNIB may be useful to map the large scale structures at the
formation epoch of the galaxies in the early Universe.

%%%%%%%%%%%%%%%%%%%%%%%%%%%%%%%%%%%%%%%%%%%%%%%%%%%%%%%%%%%%%%%%%%%%
\section{SUMMARY}

We investigated a \lya line transfer problem in the anisotropically
expanding \hone halos surrounding the first galaxies which are the
sources of \lya photons. Using the improved numerical code, we solved
the problem with 10 times larger number of test photons than that of
KK04, and examined how much the detectability of the first galaxies at
\lya lines depends on viewing angles to them. We found that the observed
profiles and the apparent luminosities of the \lya emission lines, and
the typical angular sizes of the first galaxies at \lya line strongly
depend on the inclination angle to the axis of the underlying CDM
filament. These physical quantities of the predicted \lya emission lines
of the first galaxies are compiled for various $\varepsilon$. The
direction along the filament axis is found to be more profitable to
detect the \lya emission than a direction perpendicular to the axis as
the \hone halo is more strongly decelerated against the cosmological
Hubble expansion. In the perpendicular direction to the axis, because of
the very broadened line profiles and the very diffuse surface brightness
distributions, there seems to be little hope to detect the \lya
emissions from the first galaxies with a typical \lya line luminosity of
LAEs at redshift $3-6$ of $\sim 10^{43}\ \mathrm{ergs\ s^{-1}}$. On the
other hand, along the filament axis, there is expected to be clusters of
galaxies as the sources of gravitational lensing. Based on our
estimations by using a simple lens model and a simple surface brightness
profile for the source, the high magnification with a factor of $\gtrsim
10$ is expected to the predicted \lya emissions. We conclude that this
high-magnification will allow us to detect the \lya emissions from the
first galaxies in this decade with the S/N $\gtrsim 10$ by
JWST. Moreover, in order to know their physical quantities like the SFRs
and the escape fractions of the ionizing radiation, it is essential to
understand the relation between the predicted \lya emission line profile
and the ratio of the apparent to the intrinsic \lya line luminosity.

%%%%%%%%%%%%%%%%%%%%%%%%%%%%%%%%%%%%%%%%%%%%%%%%%%%%%%%%%%%%%%%%%%%%%

\acknowledgments

We are grateful to the referee for his/her advisable comment and
excellent refereeing, which improve our paper very much in both
content and presentation.
We thank Tomonori Totani for his continuous encouragements, and Eric
Bell and Masataka Ando for useful comments.
This work is supported by the Grant-in-Aid from the Ministry of 
Education, Culture, Sports, Science and Technology (MEXT) of Japan
(16740110) and the Grant-in-Aid for the 21st Century COE
"Center for Diversity and Universality in Physics" from MEXT of Japan.

%%%%%%%%%%%%%%%%%%%%%%%%%%%%%%%%%%%%%%%%%%%%%%%%%%%%%%%%%%%%%%%%%%%%%

%%%%%%%%%%%%%%%%%%%%%%%%%%%%%%%%%%%%%%%%%%%%%%%%%%%%%%%%%%%%%%%%%%%%%

% For papers with more than 8 authors, the last name and initials of the
% first author only should be listed, followed by a comma and et al. 

%%%%%%%%%%%%%%%%%%%%%%%%%%%%%%%%%%%%%%%%%%%%%%%%%%%%%%%%%%%%%%%%%%%%%

\clearpage

%%%%%%%%%%%%%%%%%%
%  < Table 1 >   %
%%%%%%%%%%%%%%%%%%
\begin{deluxetable}{cccccccccc}
%%%%%%%%%%%%%%%%%%%%%%%%%%%
% table preamble commands %
%%%%%%%%%%%%%%%%%%%%%%%%%%%
 \tabletypesize{\scriptsize}
 \rotate
 \tablewidth{0pt}
% \tablenum{1}
 \tablecolumns{10}
 \tablecaption{Summary of the physical quantities of the predicted \lya
 line of the first galaxies. \label{tab1}}

 \tablehead{
 %%%%% 1st row headings
 \colhead{} &
 \multicolumn{2}{c}{$\lambda_\mathrm{peak}$ (\micron)\tablenotemark{a}} &
 \multicolumn{3}{c}{FWHM (\AA)} &
 \multicolumn{4}{c}{Apparent Line Luminosities\tablenotemark{b}}\\
 %%%%% 2nd row headings
 \colhead{$\varepsilon$} &
 \colhead{\phantom{ss} $\varphi_{\alpha}^\mathrm{z}\left(\lambda
 \right)$
 \phantom{s}} & 	 
 \colhead{\phantom{s}  $\varphi_{\alpha}^\mathrm{r}\left(\lambda
 \right)$
 \phantom{ss}} &	 
 \colhead{\phantom{ss} $\varphi_{\alpha}^\mathrm{z}\left(\lambda
 \right)$ 
 \phantom{s}} & 	 
 \colhead{\phantom{s}  $\varphi_{\alpha}^\mathrm{r}\left(\lambda
 \right)$
 \phantom{s}} &
 \colhead{\phantom{s} $r/z$\tablenotemark{c} \phantom{ss}} &
 \colhead{\phantom{ss} $\tilde{L}_{\alpha}^\mathrm{mean}$} \phantom{s} & 
 \colhead{\phantom{s}  $\tilde{L}_{\alpha}^\mathrm{z}$} \phantom{s} & 
 \colhead{\phantom{s}  $\tilde{L}_{\alpha}^\mathrm{r}$} \phantom{s} &
 \colhead{\phantom{s}  $\tilde{L}_{\alpha}^\mathrm{z} / 
 \tilde{L}_{\alpha}^\mathrm{r}$\phantom{s}}
 }

 \startdata
 1.0.....................................
 & 1.339 & 1.339 & 52.7 & 52.7 & 1.0 & 0.98 & 0.70 & 0.70 & 1.0 \\
 0.9.....................................
 & 1.339 & 1.340 & 68.3 & 72.0 & 1.1 & 0.97 & 0.71 & 0.62 & 1.1\\
 0.7.....................................
 & 1.340 & 1.342 & 84.8 & 115  & 1.4 & 0.96 & 0.71 & 0.42 & 1.7\\
 0.5.....................................
 & 1.342 & 1.347 & 118  & 213  & 1.8 & 0.95 & 0.66 & 0.18 & 3.7\\
 0.3.....................................
 & 1.346 & 1.363 & 190  & 506  & 2.7 & 0.92 & 0.46 & 0.01 & 43.6
 \enddata
 
 \tablenotetext{a}{
 The wavelength at which predicted \lya line profiles have their peak
 values.
 }
 \tablenotetext{b}{
 The apparent \lya line luminosities obtained by integrating the
 observed \lya line profiles with wavelength in certain ranges and by
 multiplying the integrated value by a factor of $4\pi$.
 See the text for the integral intervals adopted in this paper.
 }
 \tablenotetext{c}{
 The ratio of the FWHM of \phialphar to that of
 $\varphi_{\alpha}^\mathrm{z} (\lambda)$.
 }
\end{deluxetable}

%%%%%%%%%%%%%%%%%%%%%%%%%%%%%%%%%%%%%%%%%%%%%%%%%%%%%%%%%%%%%%%%%%%%%

\clearpage

%%%%%%%%%%%%%%%%%%
%  < Figure 1 >  %
%%%%%%%%%%%%%%%%%%

\clearpage

%%%%%%%%%%%%%%%%%%
%  < Figure 2 >  %
%%%%%%%%%%%%%%%%%%

\clearpage

%%%%%%%%%%%%%%%%%%
%  < Figure 3 >  %
%%%%%%%%%%%%%%%%%%

\clearpage

%%%%%%%%%%%%%%%%%%
%  < Figure 4 >  %
%%%%%%%%%%%%%%%%%%

\end{document}